\documentclass{article}

\usepackage[preprint,nonatbib]{nips_2018}

\usepackage{amsmath,amssymb,amsfonts}
\usepackage{algorithmic}
\usepackage{graphicx}
\usepackage{textcomp}
\usepackage{xcolor}
\usepackage{tikz}
\usetikzlibrary{patterns}

\usepackage{pgfplots}

\usepackage{listings}
\usepackage[textwidth=1.55cm,textsize=scriptsize]{todonotes}

\usepackage{array}
\newcolumntype{L}[1]{>{\raggedright\let\newline\\\arraybackslash\hspace{0pt}}m{#1}}
\newcolumntype{C}[1]{>{\centering\let\newline\\\arraybackslash\hspace{0pt}}m{#1}}
\newcolumntype{R}[1]{>{\raggedleft\let\newline\\\arraybackslash\hspace{0pt}}m{#1}}

\def\BibTeX{{\rm B\kern-.05em{\sc i\kern-.025em b}\kern-.08em
    T\kern-.1667em\lower.7ex\hbox{E}\kern-.125emX}}
\begin{document}

\title{Studying EM Pulse Effects on Superscalar Microarchitectures at ISA Level}

\author{
Julien Proy \\
INVIA \\
Meyreuil, France \\
\texttt{julien.proy@invia.fr}
\And
Karine Heydemann \\
Sorbonne Universit\'e, CNRS, LIP6 \\
Paris, France \\
\texttt{karine.heydemann@lip6.fr}
\AND
Fabien Maj\'{e}ric \\
Gemalto \\
La Ciotat, France \\
\texttt{fabien.majeric@gemalto.com}
\And
Albert Cohen \\
Inria, \'{E}cole Normale Sup\'{e}rieure, DI \\
Paris, France \\
\texttt{albert.cohen@inria.fr}
\And
Alexandre Berzati \\
INVIA \\
Meyreuil, France \\
\texttt{alexandre.berzati@invia.fr}
}

\maketitle

\begin{abstract}
In the area of physical attacks, system-on-chip (SoC) designs have not received the same level of attention as simpler micro-controllers. We try to model the behavior of secure software running on a superscalar out-of-order microprocessor typical of more complex SoC, in the presence of electromagnetic (EM) pulses. We first show that it is possible, in a black box approach, to corrupt the loop iteration count of both original and hardened versions of two sensitive loops. We propose a characterization methodology based on very simple codes, to understand and classify the fault effects at the level of the instruction set architecture (ISA). The resulting classification includes the well established instruction skip and register corruption models, as well as new effects specific to more complex processors, such as operand substitution, multiple correlated register corruptions, advanced control-flow hijacking, and combinations of all reported effects. This diversity and complexity of effects can lead to powerful attacks. The proposed methodology and fault classification at ISA level is a first step towards a more complete characterization. It is also a tool supporting the designers of software and hardware countermeasures.
	

\end{abstract}


\section{Introduction}


Physical attacks and specifically fault attacks have long been a serious threat in the world of embedded secure devices. Sensitive data are stored in devices like smart cards and passports, designed to resist such physical attacks. Tamper resistance is also part of the certification of security protocols, encryption and authentication codes, boot loaders, etc., and the devices themselves~\cite{common-criteria}.  A variety of techniques, such as electromagnetic (EM) pulses, power or clock glitches, laser beams, have been applied to a variety of devices~\cite{yuce2018}.

In an era of ubiquitous smart devices, higher integration enables complex, higher performance system-on-chip (SoC) architectures to take over simpler micro-controllers in a wide variety of secure applications. Until recently, microprocessor architecture and SoC complexity were considered strong deterrents for thwarting fault attacks~\cite{evita}. However, recent papers showed that fault attacks are also effective on such devices~\cite{vasselle, 8167704, majeric}, prompting the attention of security researchers to complex microarchitectures.

EM fault injection does not require specific chip preparation and offers an interesting trade-off between affordability and precision. 
It has been proven effective on simple microarchitectures~\cite{Moro_2013, cryptoeprint:2012:123} and features relatively well understood fault models~\cite{Moro_2013, 10.1007/978-3-319-31271-2_7}. SoCs and complex architectures have not received the same level of attention yet. A handful of recent studies targeting complex SoC reported feasible exploitation~\cite{majeric}. However, the achievable effects remain poorly understood; characterizing the effects of EM pulses on complex microarchitectures is a necessary step to enable the design of effective software and hardware countermeasures.

This paper analyzes the behavior of secure software running on a superscalar out-of-order microprocessor in the presence of EM pulses. Following a black box approach, we successfully corrupt the loop iteration count of vanilla as well as hardened versions of two sensitive loops running on an ARM Cortex-A9 processor. We propose a step-by-step methodology using very simple codes to understand and classify the fault effects at the level of the instruction set architecture (ISA).\footnote{The ISA is the lowest level available to a programmer implementing secure applications.} The well established instruction skip and register corruption models are part of this classification, together with new effects specific to more complex processors. The latter include operand substitution, multiple correlated register corruptions, advanced control-flow hijacking, and combinations of all reported effects. This diversity and complexity of effects can lead to powerful attacks. The proposed ISA-level classification and analysis on secure software is a first step towards a more complete characterization of EM-induced faults.



The paper is organized as follows. Section~\ref{sec:related} discusses related work. Section~\ref{sec:setup} describes the hardware and software setup. Serving as a motivation for the whole study, Section~\ref{sec:motiv} conducts a sensitivity analysis of vanilla and hardened versions of representative loops. Section~\ref{sec:carac} presents a step-by-step methodology and derives an initial fault model classification. We further extend these models to interpret the faults observed on the sensitive loops in Section~\ref{sec:loops}, wrapping up a first characterization. Section~\ref{sec:ccl} concludes with directions towards a more complete characterization.

\section{Related work}\label{sec:related}

Since the pioneering exploit of a CRT-RSA implementation~\cite{Boneh1997}, fault attacks targeting cryptographic applications have been under the spotlight of security research and new studies are published every year~\cite{yuce2018}.  Such fault attacks are either described at an algorithmic level~\cite{cryptoeprint:2012:123, cryptoeprint:2002:073} or target a specific implementation~\cite{DehbaouiMMDT13,cryptoeprint:2011:388}.
They rely on the attacker's ability to inject and exploit fault effects on a sensitive application: e.g., a skipped instruction or corrupted variable. As a result, most studies considered the injection or exploitation dimensions, while the characterization of possible fault effects received less attention. This is particularly true of modeling efforts targeting the hardware/software interface.


Moro et al.\ proposed a first ISA-level characterization of EM fault injection on a 32-bit ARM Cortex-M3 processor~\cite{Moro_2013}. Observing that EM fault injections can alter transfers from the Flash memory, they model these effects as load value corruption or instruction replacement at ISA level. Also, 25\% of these instruction replacements happen to be equivalent to skipping an instruction.
Riviere et al.\ reported that due to the presence of an instruction cache or prefetch buffer, some instructions (previously fetched) can be replayed as a consequence of EM fault injection on an ARM Cortex-M4 processor~\cite{riviere:hal-01208378}.
Yuce et al.~\cite{Schaumont_fdtc2016} targeted a 7-stage-pipeline 32-bit LEON3 processor using voltage glitch injection, which can lead to the corruption of up to 5 consecutive instructions---each one in a different pipeline stage.  

Dureuil et al.~\cite{10.1007/978-3-319-31271-2_7} proposed to infer a probabilistic fault model from a series of iterative fault injection campaigns running dedicated test codes. This model is the basis for the robustness evaluation of an application using a simulator. While the accuracy of the inferred probabilistic fault model is highly dependent on the first set of experiments, this work highlights the importance of crafting adequate test codes to analyse fault effects.


Kelly et al.\ proposed a methodology to characterize a fault model from laser injections on an 8-bit AVR micro-controller~\cite{7951802}. They performed a whole chip scan running test codes consisting of a single instruction from different classes. They mapped every instruction (class) to one or more sensitive areas on the chip and associated it with different observable effects.
A deeper investigation revealed that laser beam injection on a selected sensitive area enables to skip an instruction. This methodology shows the importance to use specific test codes to determine fault effects at ISA level, yet the authors did not propose a classification of all observed effects. 

 



The injection of faults on more complex SoC has recently raised the interest of security researchers.
Attacks targeting an ARM Cortex-A9 platform---typical of early smartphones--- attempted priviledge escalation in Linux using power glitches~\cite{8167704} and to bypass secure boot protections using laser injection~\cite{vasselle}.
Majeric et al.~\cite{majeric} sucessfully injected EM-induced faults targeting a hardware AES implementation on a similar SoC.  These early studies demonstrate the feasibility to target a specific SoC location without altering the behavior of the full chip.
All these works report on the feasibility of exploiting fault injections, but their effects at ISA level are not yet understood. The characterization of fault effects on complex microarchitectures is yet to come, in particular the effects of EM fault injection.

\section{Experimental setup}
\label{sec:setup}

This section details the fault injection setup including the targeted hardware device and the software environment supporting the attack campaigns and analyses.

\subsection{Device under test}

All attacks have been realized on a widespread SoC typical of automotive and Internet of Things (IoT) applications. It consists of a dual-core 32-bit ARM Cortex-A9 MPCore on CMOS 40nm technology implementing the ARMv7-a ISA and clocked at up to 1 GHz. We clocked the chip at 80MHz for observability purposes.

The following features of the Cortex-A9 microarchitecture stand out from previous studies of simpler in-order processors such as secure element micro-controllers~\cite{Schaumont_fdtc2016, Moro_2013}:
\begin{itemize}
\item 8-stage variable-latency pipeline;
\item superscalar, dual-issue instruction decoder;
\item register renaming with out-of-order write-back;
\item branch predictor (with a branch target buffer);
\item 64B loop buffer bypassing the instruction cache.
\end{itemize}

The memory hierarchy features separate level~1 instruction and data caches of 32KB each, a unified level~2 cache of 256KB, along with several external memory interfaces (DDR3, NAND-Flash and NOR-Flash). The SoC is embedded in a development board suitable for instrumentation on the test bench, including a serial port (UART) and GPIO signals.

\subsection{Attack setup}

\begin{figure}[!ht]
	\centering
	\includegraphics[width=0.8\columnwidth]{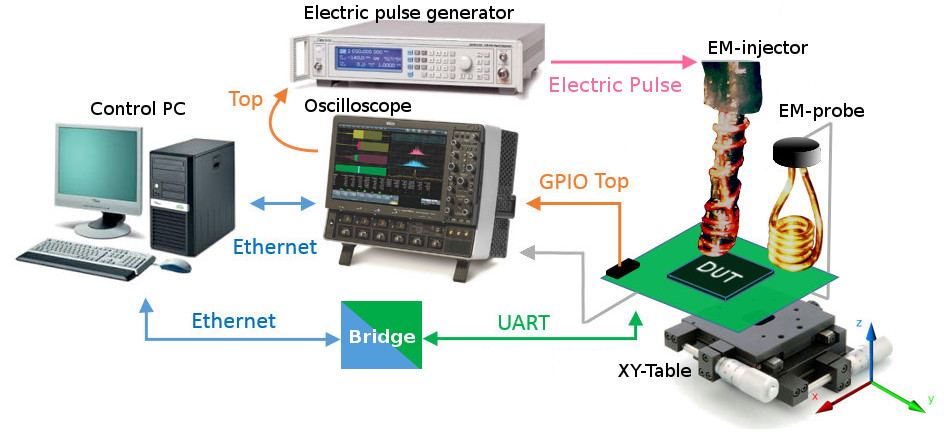}
	\caption{Diagram of the automated EM injection platform}
	\label{fig:EMbench}
\end{figure}

Electromagnetic fields (EM) are a suitable physical quantity to observe or disturb a running processor. 
We used the EM injection bench presented in \figurename~\ref{fig:EMbench}. It is composed of an automated XY-table where the target board is placed, a generator capable of injecting pulses from 6ns to 150ns and up to $\pm$ 400V, an oscilloscope, an EM probe and an EM injector. The EM probe is a coil of copper used to monitor and time the moment of the injection very precisely, and to observe the effects. The EM injector is also a coil of copper, reeled around a piece of ferrite to focus the EM field. The board is connected through a serial port to a control PC to communicate data for analysis. Every component is connected to the control PC which synchronizes the necessary actions after an initial GPIO trigger.

\begin{figure}[!ht]
	\centering
	\begin{tabular}{C{4cm} C{4cm}}
		{\includegraphics[scale=0.3]{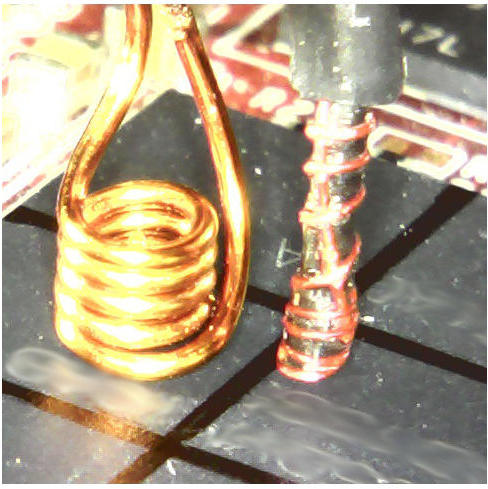}} & {\includegraphics[scale=0.3]{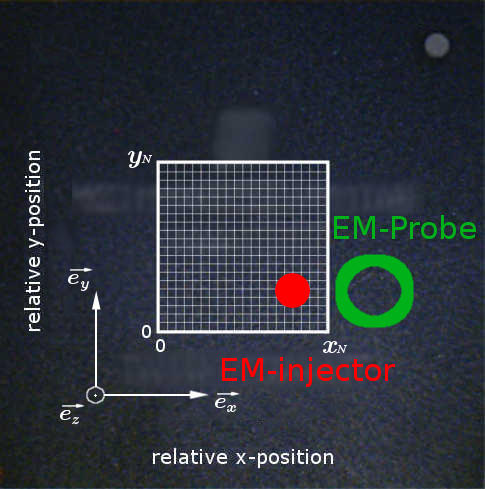}}\\
	\end{tabular}
	\caption{EM probe and EM injector (left) as well as their positioning over the SoC surface (right); the axis represents the SoC package}
	\label{fig:SoCandProbes}
\end{figure}

Multiple parameters must be set when injecting a fault such as spatial location, temporal location, injection voltage, pulse polarity and duration.
The parameters space is too large to be exhaustively tested. One has to focus on most meaningful parameters~\cite{riviere:hal-01208378}.
Our goal was to obtain a successful setup (spatial location, voltage and duration of injection) that injects exploitable faults with high probability.  Following a trial and error scheme~\cite{10.1007/978-3-319-31271-2_7}, we selected the sensitive location highlighted in \figurename~\ref{fig:SoCandProbes}, a duration of 6ns and a pulse voltage of 310V.
Lower voltage reduces the occurrence of successful faults while higher voltage mutes the SoC more frequently.
Overall, the \emph{timing offset} to inject the fault after the initial GPIO trigger is the only parameter varying during our experiments. It lets us target different instructions as the processor runs a given test code.
In the following, an \emph{attack campaign} refers to series of experiments on the same test code, each experiment consisting of a single EM pulse, and varying the timing offset of injection. 

\subsection{Software setup}

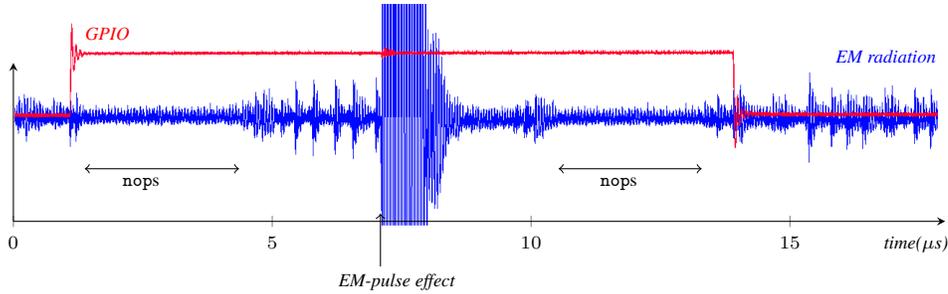
\begin{figure}[ht]
\begin{tikzpicture}

\draw (6,1.4) node[inner sep=0] {\includegraphics[width=0.9\columnwidth]{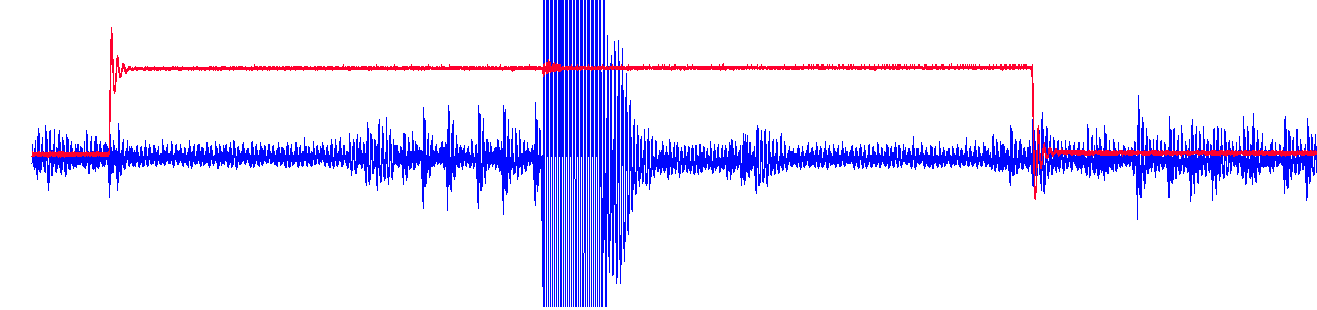}};

\begin{axis}[
y=30pt,
width = \columnwidth,
xmin=0,
xmax=18,
ymin=-1,
ymax=1,
ytick style={draw=none},
axis lines = left,
xtick={0,5,10,15},
x tick label style={font=\scriptsize},
yticklabels={,,},
]
\end{axis}

\draw [<->] (0.95,0.7) -- (3,0.7);
\draw (1.7,0.5) node {\scriptsize{\texttt{nop}s}};

\draw [<->] (7.25,0.7) -- (9.15,0.7);
\draw (8.05,0.5) node {\scriptsize{\texttt{nop}s}};

\draw (12,-0.3) node {\scriptsize{\textit{time($\mu$s)}}};

\draw (1.25,2.5) node {\textit{\scriptsize{\textcolor{red}{GPIO}}}};
\draw (11.6,2.2) node {\textit{\scriptsize{\textcolor{blue}{EM radiation}}}};

\draw [->] (4.88,-0.6) -- (4.88,0.1);
\draw (5.1,-0.8) node {\textit{\scriptsize{EM-pulse effect}}};
\end{tikzpicture}
\caption{EM radiation emanating from the chip while injecting a fault during the execution of a loop}
\label{fig:curvefab}
\end{figure}

We derive the most appropriate instant to inject an EM disturbance from the EM probe measurements. First of all, to ease the detection of the targeted code's pattern in the EM trace, we inserted a sequence of 200 \texttt{nop}s before and after the targeted region in the binary code. These sequences exhibit much lower EM emissions that are easily detected on the trace.
\figurename~\ref{fig:curvefab} shows the EM trace of an execution of a loop where the pulse has been injected at the fifth iteration. 
In \figurename~\ref{fig:curvefab}, we can observe the different loop iterations (small peaks) as well as \texttt{nop} sequences (flat parts), and the fault injection itself (high oversized and truncated peak). 


Inserting \texttt{nop} sequences and coupling those with a GPIO trigger help to determine an interval of maximal timing offsets covering a significant part of the running trace of the target code. E.g., multiple iterations for a loop.
An attack campaign steps through that interval, injecting one pulse at a precise timing offset; we selected steps of 5ns to be shorter than CPU clock period.
Besides, the execution of any test code ends with a sequence of instructions to store the contents of general-purpose registers \texttt{r0} to \texttt{r13} and to copy its operating memory areas into an \emph{output buffer}. This output buffer is sent to the PC for analysis.

A dry run without any injection allows to collect reference values in the output buffer. Every subsequent run is subject to an EM fault injection. Comparing the output buffer with reference values is called \emph{result analysis} in the following.
We made the following implementation choices to facilitate result analysis:
\begin{itemize}
\item the general-purpose registers are initialized with distinct low Hamming weight values, to help narrowing down the analysis in case of (multiple) register corruption;
\item similarly, the contents of operating memory areas is initialized with distinct ``remarkable'' values;
\item the test codes are developed in C and compiled with clang/llvm version 6.0 at optimization level -O2; the generated assembly code has been manually rewritten to use all available registers (register allocators typically aims to spare registers) to better track the fault effects on intermediate computations until the end of the attacked code region.
\end{itemize}

For every fault injection, different types of results can be observed. 
They are classified and assigned to different groups among:
\begin{itemize}
\item \textit{no fault}: the output buffer content exactly matches the reference one: the fault has no visible effect;
\item \textit{successful fault}: the output buffer contents does not match the reference one;
\item \textit{mute}---the board did not reply to the command; its status is undefined (the board needs a hard reset).
\end{itemize}

For each successful fault, the received values may be analyzed to help identify what happened at ISA level.

\section{Preliminary fault sensitivity analysis}
\label{sec:motiv}

Attacks in the cryptography~\cite{DehbaouiMMDT13,EspitauFGT18} and systems~\cite{nashimoto_buffer_2016} literature often target loops, aiming for early or deferred loop exit. We would first like to assess the practicality of such control flow disruption on sensitive loops.

\subsection{Loop benchmarks}

We selected two loops for their representativity of sensitive code and small enough to make the analysis of fault effects tractable:
\begin{itemize}
\item the \textit{memcpy}-like function in Listing~\ref{loop1-c} is typical of firmware updates subject to buffer overflow attacks~\cite{nashimoto_buffer_2016}; instead of copying, values in the source and destination buffers are added to ease result analysis;
\item the \textit{memcmp}-like function with early exit in Listing~\ref{loop2-c} resembles authentication schemes such as PIN verification~\cite{fissc};
again, to facilitate analysis, the outcome of the comparison is stored in a destination buffer at every iteration.
\end{itemize}

\begin{figure}[ht]
\hfill
\begin{minipage}{0.42\linewidth}
\begin{center}
\begin{lstlisting}[language=c, caption=Simple loop, frame = single, label=loop1-c]
 for(k=0; k<n; k++) {
   dst[k] += src[k];
 }


\end{lstlisting}
\end{center}
\end{minipage}\hfill
\begin{minipage}{0.52\linewidth}
\begin{center}
\begin{lstlisting}[language=c, caption=Multiple exit loop, frame = single, label=loop2-c]
 for (k=0; k<n; k++) {
    dst[k] = (src1[k] != src2[k]);  
    if (dst[k])
      break;
 }
\end{lstlisting}
\end{center}
\end{minipage}\hfill
\vspace{-0.2cm}
\end{figure}

The second loop is structurally more complex than the first one: it has two different exits. Moreover, its early exit condition depends on a comparison between data read from memory whereas the loop exit of the first loop only depends on a monotonically increasing counter.

\subsection{First fault injection campaign}

We conducted an attack campaign on both loops, stepping pulse injections through all instructions executed in a given iteration.
Table~\ref{tab:nosec-nf} shows the results.
\begin{table}[ht]
\caption{Classification of EM pulse results}\label{tab:nosec-nf}
\centering
\begin{tabular}{c|c c c}
  \hline
  Code & No Fault & Mute & Successful faults \\
  \hline
  \textit{loop1} & 12663 (93.0\%) & 403 (2.9\%) & 555 (4.1\%)\\
  \textit{loop2} & 14287 (95.0\%) & 341 (2.4\%) & 372 (2.6\%)\\
  \hline
\end{tabular}
\end{table}

The fraction of successful faults is consistent with existing results on simpler devices \cite{uboot}. Successful faults can be further divided into two classes: a \emph{harmful fault} led to corrupted output values and taking the wrong exit (incorrect number of iterations or taking the wrong exit in the second loop); a \emph{harmless fault} where output are not nominal but still indicate a correct number of loop iterations and taken exit branch. Table~\ref{tab:nosec-sf} shows the proportion of successful faults that are classified as harmful or harmless: 15\% of the successful faults on the first loop break the security property and up to 80\% of the successful faults on the second loop. Since the latter involves more registers and instructions in its exit conditions, its attack surface is higher, explaining the higher number of successful faults. 
These campaigns demonstrate our injection setup's effectiveness on simple loops on a complex microarchitecture.

\begin{table}[ht]
\caption{Breakdown of successful faults}\label{tab:nosec-sf}
\centering
\begin{tabular}{c|c c c}
  \hline
  Code & Harmful faults & Harmless faults \\
  \hline
  \textit{loop1} & 87 (15.7\%) & 468 (84.3\%) \\
  \textit{loop2} & 299 (80.4\%) & 73 (19.6\%) \\
  \hline
\end{tabular}
\end{table}

Our first attempt to explain these successful faults is to consider existing fault models, such as instruction skip \cite{Moro_2013} and register corruption \cite{1580506, Ordas2014, Ordas2015}.
We analyzed the faulted contents of the output buffer and observed that 8\% (resp.\ 14\%) of the faulty behaviors can be explained by an instruction skip (resp.\ register corruption).
The remaining 78\% successful faults cannot directly be explained without further investigation.

In practice, sensitive codes are often protected against instruction skip and register corruption~\cite{plaf, Reis2005, thierno, DeKeulenaer2015, Lalande2014}. For this reason, the following section studies the feasibility of attacking hardened codes.

\subsection{Fault injection campaign on hardened code}

We selected two hardening schemes from the literature, one dedicated to loops~\cite{plaf} and a general-purpose scheme~\cite{Reis2005}.

The first scheme replicates the loop exit condition and the slice of instructions involved in its computation. Both conditions are compared at each iteration. Any fault impacting one of the computations is detected and leads to an error handler. This countermeasure is designed to resist under the instruction skip and register corruption fault models.

The second scheme, called \emph{SWIFT}, is based on a duplication of all instructions. This scheme also introduces a signature-based control-flow integrity mechanism: every basic block has its own signature; two variables track signature updates at branches using both original and duplicated predicates.
This scheme is designed to detect control-flow hijacking and any register corruption resulting from one \emph{Single Event Upset}~\cite{1580506} where a single bit is flipped.

These countermeasures are designed to be target independent. 
The loop hardening scheme was validated on a simple micro-controller, the ARM Cortex-M3~\cite{plaf}; implemented in the llvm middle-end, it is immediately applicable to more complex processors with similar ISA (ARMv7-m vs.\ ARMv7-a).
SWIFT was originally designed for IA64, taking advantage of its instruction predication and wide-issue logic. We adapted it to the ARMv7 assembly language.
Since both schemes behave very similarly on the simple control flow of the \textit{loop1} benchmark, we report the SWIFT results on the \textit{loop2} benchmark only.

In the following, \textit{loop1-sec} and \textit{loop2-sec} refer to the application of the loop hardening scheme to both loop benchmarks, whereas \textit{loop2-swift} corresponds to the second loop hardened with SWIFT.

We conducted attack campaigns on these hardened loops. The results include an additional class called \emph{Detected fault}, counting EM pulses resulting into a detected fault with a corrupted output buffer.
 
\begin{figure}
\centering
\begin{tikzpicture}
	\begin{axis}[
	xbar stacked,
	bar width=9pt,	
	y=22pt,
	width = 22em,
	ymajorgrids = true,
	xmin=0,
	xmax=100,
	ymin=0.2,
	ymax=1.8,
	xtick={0,20,40,60,80,100},
	xticklabels={0,20,40,60,80,100\%},
	x tick label style={font=\footnotesize},
	legend style={at={(0.46,-0.75)}, anchor=south, draw=none, legend columns=-1, column sep=1ex, font=\footnotesize},
	yticklabels={
		loop1-sec,
		loop2-sec,
		loop2-swift,
	},
	ytick=data,
	y tick label style={font=\scriptsize},
	]
	
	\addplot[fill=white] plot coordinates {
		(52.8, 1.5)(22.2, 1)(50.9, 0.5)
	};
	\addplot[fill=black!20] plot coordinates {
		(33.6, 1.5)(69.3, 1)(19.4, 0.5)
	};
	\addplot[fill=black!40] plot coordinates {
		(13.6, 1.5)(08.5, 1)(29.7, 0.5)
	};
	\legend{Harmless fault, Detected fault, Harmful fault}
	\end{axis}
\end{tikzpicture}
\caption{Successful fault breakdown on hardened loops}\label{fig:prop}
\end{figure}
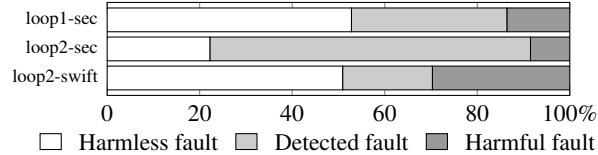

\figurename~\ref{fig:prop} shows interesting results.
Hardening the code significantly reduces the probability of a successful attack, and this is the case for both countermeasures.
The countermeasures do detect some of the faults, this is consistent with our preliminary analysis that some injections led to skipping an instruction or corrupting a register.
However, as numerous faults are not detected, these results also suggest that EM pulses may induce complex effects at software level that are not caught by the countermeasures. Further fault characterization is required to better understand the achievable effects visible at ISA level.

\section{Characterizing and modeling faults}
\label{sec:carac}

To characterize the different fault effects, let us step back from loop benchmarks and switch to a set of much simpler (synthetic) codes. The simpler codes are designed to exercise a narrower set of microarchitectural elements (registers, functional units, data path). The attack campaign considers codes of increasing complexity and microarchitectural footprint. This step by step methodology facilitates the isolation of (faulty) effects; it also enables designing subsequent experiments from the previously observed effects.






\subsection{Sequence of \texttt{nop}}

Without prior knowledge on the effect of electromagnetic injection on this complex SoC, we started from the most basic example consisting of a linear sequence of \texttt{nop} instructions.

On a campaign of 3000 runs, we observed 142 \textit{mute}s but no successful fault.
It appears that, at ISA level, neither registers nor instructions have been corrupted. This was somewhat surprising to us given the observed register corruption on the loop benchmarks.
One may immediately deduce that there is no general instruction replacement going on: any replacement, if it occurs, has no effect on the architectural registers. In particular, this experiment rules out any of thin air instruction replacement on or before the decode stage.

By construction, it is not possible to detect instruction skip on such a code, which brings us to the following experiment.

\subsection{Single counter incrementation}

Our second test example consists of a sequence of \texttt{add r0, r0, \#1} instructions.
By incrementing only one register, we attempt to correlate corruption effects with instruction operands.

Among all successful faults, only \texttt{r0} has been corrupted. This is an important result consistent with the previous experiment on a sequence of \texttt{nop}s. 
It also indicates that the microarchitecture does not implement \texttt{nop} as an instruction involving one or more register operands (e.g., \texttt{add r0, r0, \#0}).

We also analyzed the faulted values observed in \texttt{r0}. Most of these correspond to twice the expected (unfaulted) value plus a small negative offset\footnote{Small relative to the initial value of \texttt{r0}, see Section~\ref{sec:setup}.}. Occasionally three or four times. This could be explained by the selection of an architectural or bypass logic register instead of the expected constant \texttt{\#1}. The offset is proportional to the number of instructions remaining to be executed after the fault affecting \texttt{r0}. We could confirm this timing correlation on the EM trace. The proportionality is faithfully observed except in very rare cases where the offset is smaller; this suggests (rare) additional effects such as instruction skip and bit flips. At ISA level, it can be modeled as the replacement of one or more instructions \texttt{add r0, r0, \#1} by \texttt{add r0, r0, r0}.
Given the selectivity of the observed values, we can also rule out a functional unit replacement,
e.g., \texttt{sll r0, r0, \#1} would be an alternative replacement but such an explanation would authorize replacements with more arithmetic and logic operations leading to a much wider distribution of observed values.

We then studied the sensitivity to the register operand number. Replacing \texttt{r0} with \texttt{r5} in the sequence of additions showed that only \texttt{r5} was faulted, which is consistent with the previous findings. Yet the faulted values now behave differently: we now observe values consistent with a majority of replacement with \texttt{add r5, r5, r0} and a minority with \texttt{add r5, r5, r5}! This motivates further study of the influence of the register operand number in the fault effects, which will be the purpose of the following experiment.

So far, 96\% of these two experiments can be explained with the replacement of one operand, but the exact nature of this replacement remains obscure at this point. Many microarchitectural effects could explain the observed results: corruption may take place in the registers themselves, in selectors/multiplexers addressing architectural or physical register banks, or the bypass logic. We are also not sure how many registers and/or instructions have been corrupted, but several observations hint at multiple faults, possibly correlated/coupled..

\subsection{Multiple counter incrementation}
The next experiment runs a sequence of \texttt{add rx, rx, Cx} instructions where \texttt{Cx} is a different constant for every register, with \texttt{x} ranging from 0 to 9.
This configuration has four advantages.
First of all, since the sequence features additions only, it mitigates complexity.
Second, it touches multiple registers, but none of these induces any dependence. Third, the distance between two instructions modifying the same register is 10 instructions, limiting time-dependent effects associated with out-of-order execution or pipelining.
Finally, every register is incremented using a different immediate constant (\texttt{Cx}). We used prime numbers to reduce chances of collision in corrupted values.

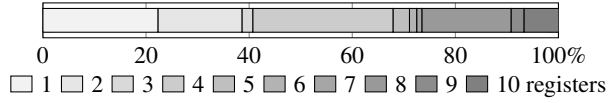
\begin{figure}
\centering
\begin{tikzpicture}
\begin{axis}[
	xbar stacked,
	width=24em,
	bar width=9pt,
	y=22pt,
	xmin=0,
	xmax=100,
	xtick={0,20,40,60,80,100},
	xticklabels={0,20,40,60,80,100\%},
	x tick label style={font=\footnotesize},
	ymajorgrids = true,
	ymin=0.7,
	ymax=1.3,
	legend style={at={(0.52,-2.0)}, anchor=south, draw=none, legend columns=-1, column sep=0.25ex, font=\footnotesize},
	yticklabels={},
]
\addplot[fill=gray!10] plot coordinates {(22.3, 1)};
\addplot[fill=gray!20] plot coordinates {(16.3, 1)};
\addplot[fill=gray!30] plot coordinates {(02.1, 1)};
\addplot[fill=gray!40] plot coordinates {(27.2, 1)};
\addplot[fill=gray!50] plot coordinates {(03.2, 1)};
\addplot[fill=gray!60] plot coordinates {(01.4, 1)};
\addplot[fill=gray!70] plot coordinates {(01.0, 1)};
\addplot[fill=gray!80] plot coordinates {(17.3, 1)};
\addplot[fill=gray!90] plot coordinates {(02.5, 1)};
\addplot[fill=gray!100] plot coordinates {(06.8, 1)};
\legend{1,2,3,4,5,6,7,8,9,10 registers}
\end{axis}
\end{tikzpicture}
\caption{Distribution of the number of corrupted registers per successful injection}\label{fig:numregs}
\vspace{-0.2cm}
\end{figure}

This campaign hints at the number of instructions or registers that may be impacted by a single injection, as well as the potential interaction between simultaneously executed instructions. And our observations indicate that many registers can be corrupted by a single fault. As previously observed, only registers occurring as instruction operands are corrupted, and we also observe that all registers have the same sensitivity to corruption or to be used as a source operand in a replacement (resulting in the corruption of another register).
\figurename~\ref{fig:numregs} shows the distribution of the number of corrupted registers impacted by one single injection.
Multiple register corruption is more likely to occur: the average number of corrupted registers per successful fault is $3.8$ and all alive registers can be corrupted at once.

The analysis is already delicate, even on such a simple code. Yet some behaviors can still be explained thanks to the setup of initial register values.
We observed that some registers received the values expected for other ones. At ISA level, it can be modeled as the replacement of \texttt{add ry, ry, Cy} by \texttt{add ry, rx, Cx} where \texttt{rx + Cx} is the value expected for another register. Some corrupted values can also be explained by an operand replacement. All these observed faults still seem to come from the selection of an architectural or bypass logic register instead of expected operands.

Some corrupted registers also derive from in-flight values (from the register itself or another one) with several flipped bits. Also, some corrupted register contents match precisely the reset of their 16 most-significant bits.
The experiment highlights another frequent effect resulting into a double register corruption.  Considering the two instructions \texttt{add r2, r2, \#5} and \texttt{add r3, r3, \#7}, we observed correlated corruptions of the form
\begin{center}
	$\textit{faulted\_value}(\texttt{r2}) - \textit{expected\_value}(\texttt{r2}) = -5$ \\
	$\textit{faulted\_value}(\texttt{r3}) - \textit{expected\_value}(\texttt{r3}) = 7$
\end{center}
These can be modeled by one instruction skip (here \texttt{add r2, r2, \#5)} and the replay of an instruction (\texttt{add r3, r3, \#7}). The frequency of this effect targeting different registers with different values rules out another cause such as bit flips. 
Some groups of corrupted registers do not belong to consecutive instructions in the program order; this is most likely due to superscalar out-of-order execution.

\subsection{Isolating effects}

In the previous attack campaign, multiple operands are often corrupted simultaneously across one or more instructions.
The injection pulse duration is really short---6ns for a 12ns clock period---and does not directly explain this behavior.
We thus suspected side effects due to pipelining where a single glitch can influence multiple instructions~\cite{Schaumont_fdtc2016}.
In an attempt to reduce the number of simultaneous corruptions, we replayed the last campaign inserting one or several \texttt{nop} instructions between every addition.
The average number of simultaneously corrupted registers effectively went down from $3.8$ to $3.2$ with 1 \texttt{nop}, $1.8$ with 2 \texttt{nop}s, $1.7$ with 4 \texttt{nop}s, and to $1.1$ with 50 \texttt{nop}s.
These experiments confirm that a single fault may affect several in-flight instructions.

\subsection{Fault models and classification}

To summarize our findings, we observed and analyzed different fault effects and highlighted different of their properties:
a register not used by the processor has a very low probability of being corrupted;
all registers seem to have the same sensitivity despite a fixed EM-pulse spatial position;
registers value can be directly corrupted (multiple bit-flips or most-significant half-word reset);
all alive registers can be corrupted at once and inserting \texttt{nop} instructions between every instruction reduces the number of corrupted registers.
Finally, instructions can be corrupted in multiple ways. They can either be skipped (no write, no operation) or replayed, or their operands can be substituted mainly by operands from other instructions being processed. 

These are general observations and analyzed behaviors for the fault injection setup.
Before confronting and applying these findings to complex test codes, we further classify the results of a single injection according to the type of effect observable at ISA level:
\begin{itemize}
\item \textit{Instruction skip}: one instruction is skipped.
\item \textit{Register most-significant half-word (mshw) reset}: the corrupted value corresponds to a reset of the 16 most-significant bits of the expected value.
\item \textit{Register corruption}: the corrupted value is either derived from another existing one (bit flips) or does not seem correlated to any in-flight value.
\item \textit{Source operand substitution}: an immediate value or a source register of an instruction is not the expected one; the corrupted value usually comes from the immediate or source register operand of another instruction.
\end{itemize}

In most cases, results cannot be explained directly by a single one of the four previous groups. Instead, a combination of these effects is needed, often with strongly coupled \emph{composite effects}:
  \begin{itemize} 
  \item Combined skip of an instruction and replay of a previously executed one.
  \item Combined register corruptions leading to correlated corrupted values.
  \item Repeated occurrences of the same fault effects, such as instruction skip or register corruption with or without correlated values.
  \item Sometimes, \textit{mixed faults} that correspond to two or more distinct of the above, without any apparent correlation.
 \end{itemize}

Moreover, the instructions impacted in these fault effect do not necessary form consecutive intervals due to out-of-order execution.
Also, the occurrence rate of all these fault patterns are not equal. The precise effect of one injection is known to be difficult to predict.  However, we observed several times some faulty outputs with the same values. Hence, the probability to reproduce some faults is far from negligible.

Based on these analysis and classification, we can revisit the yet unexplained results from more complex loop examples. This is the purpose of the next section.



\section{Looping back}
\label{sec:loops}

A major difference between the previous test codes and the more complex loops resides in the instruction mix.
In addition to arithmetic operations, loops contain memory accesses (load/store) and branches, and access more diverse data coming from memory.
Also, the results are more difficult to analyze due to the propagation of the fault effects until the contents of the output buffer can be retrieved.

\subsection{Loop analysis w.r.t.\ fault classification}
\label{sec:classification}

To analyze results from successful faults on loop test codes, we first considered the fault effect classification established in the previous section. As a faulted contents of the output buffer can sometimes be explained with more than one single ISA-level effect, we prioritized the models in reverse order of their (apparent) complexity: we first look for a skipped instruction, and subsequently for operand substitution, register corruption, composite faults, and eventually mixed models. We managed to classify a significant part of the results, but some effects remain unexplained when considering these fault models only.

Some values loaded from memory areas have been corrupted while source memory remains unaltered. These values follow two patterns: either similar to alive register contents with several flipped bits or values apparently not correlated to any available values but repeated over injections. The latter ones can be explained by values from other memory areas and thus be related to an operand substitution in the load instruction. The former pattern can be modeled by a register corruption as already observed. However, we cannot ensure that it does not come from a corruption of the memory transfer itself. Moreover, we encountered this case several times highlighting a sensitivity of load instruction. We call this a \emph{load corruption} effect. 


\begin{figure*}
\resizebox{\linewidth}{!}{
\begin{tikzpicture}
\begin{axis}[
xbar stacked,
width=43em,
bar width=9pt,
y=22pt,
ymajorgrids = true,
xmin=0,
xmax=100,
ymin=0.2,
ymax=3.3,
xtick={0,10,20,30,40,50,60,70,80,90,100},
xticklabels={0,10,20,30,40,50,60,70,80,90,100\%},
x tick label style={font=\footnotesize},
yticklabels={
	loop2-swift+sec (2074),
	loop2-swift (1615),
	loop2-sec (702),
	loop2 (372),
	loop1-sec (475),
	loop1 (551),
},
y tick label style={font=\scriptsize},
legend style={at={(0.41,-0.45)}, anchor=south, draw=none, legend columns=-1, column sep=0.4ex, font=\footnotesize},
ytick=data
]
\addplot[fill=black!65] plot coordinates {
	(52.0, 0.5)(20.7, 1)(15.5, 1.5)(50.0, 2.0)(14.8, 2.5)(26.5, 3.0)
};
\addplot[fill=black!40] plot coordinates {
	(13.2, 0.5)(17.6, 1)(08.4, 1.5)(05.1, 2.0)(12.7, 2.5)(00.9, 3.0)
};
\addplot[pattern=north west lines] plot coordinates {
	(09.5, 0.5)(08.8, 1)(11.4, 1.5)(17.5, 2.0)(16.9, 2.5)(10.2, 3.0)
};
\addplot[pattern=crosshatch] plot coordinates {
	(00.0, 0.5)(00.1, 1)(01.4, 1.5)(01.1, 2.0)(00.2, 2.5)(00.0, 3.0)
};
\addplot[pattern=dots] plot coordinates {
	(15.9, 0.5)(24.1, 1)(34.1, 1.5)(17.5, 2.0)(13.9, 2.5)(18.0, 3.0)
};
\addplot[pattern=north east lines] plot coordinates {
	(01.5, 0.5)(01.2, 1)(09.1, 1.5)(00.0, 2.0)(18.8, 2.5)(00.0, 3.0)
};

\addplot[fill=black!65] plot coordinates {
	(02.1, 0.5)(02.3, 1)(07.6, 1.5)(00.5, 2.0)(01.9, 2.5)(06.5, 3.0)
};
\addplot[fill=black!40] plot coordinates {
	(00.2, 0.5)(02.7, 1)(01.1, 1.5)(00.8, 2.0)(00.0, 2.5)(00.2, 3.0)
};
\addplot[pattern=north west lines] plot coordinates {
	(00.4, 0.5)(00.7, 1)(01.7, 1.5)(00.0, 2.0)(02.3, 2.5)(17.4, 3.0)
};
\addplot[pattern=crosshatch] plot coordinates {
	(00.0, 0.5)(00.0, 1)(00.0, 1.5)(00.0, 2.0)(00.0, 2.5)(03.4, 3.0)
};
\addplot[pattern=dots] plot coordinates {
	(00.6, 0.5)(00.1, 1)(00.3, 1.5)(00.0, 2.0)(00.0, 2.5)(00.0, 3.0)
};
\addplot[fill=black!15] plot coordinates {
	(04.6, 0.5)(21.7, 1)(09.4, 1.5)(07.5, 2.0)(18.4, 2.5)(20.0, 3.0)
};

\legend{instruction skip, register corruption, operand replacement, \textit{mshw} reset, load corruption, magic-edge,,,,,, mixed faults}
\end{axis}

\draw [line width=1.2pt, black] (7.54,2.75) -- (7.54,1.975) -- (10.45,1.975) -- (10.45,1.585) -- (12.32,1.585) -- (12.32,1.20) -- (10.83,1.20) -- (10.83,0.81) -- (9.8,0.81) -- (9.8,0.42) -- (12.45,0.42) -- (12.45,0);
\draw [line width=1.2pt, black] (7.42,2.75) -- (7.66,2.75);
\draw (4.5,2.60) node {\footnotesize{\textit{single faults}}};
\draw (12,2.60) node {\footnotesize{\textit{composite faults}}};

\end{tikzpicture}
}
\caption{Fault distribution according to the different models and the multiplicity of faults for each loop code (with the total number of \textit{successful faults} for each code)}\label{fig:models}
\end{figure*}
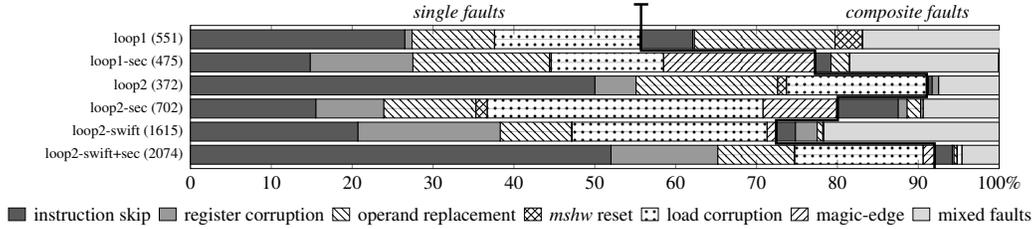


The second of these new effects is much more original and is best understood at the level of the control-flow graph (CFG) of the function. Some injections on hardened versions can only be explained by a jump from the end of a basic block (linear sequence of instructions without a branch) \emph{directly to the beginning of another one}. When the target block is an illegal destination block in the CFG, we name this effect a \emph{magic edge}.  We did not observe any random jumps to the middle of basic blocks.
This behavior hints at the corruption of the branch target buffer or another branch prediction mechanism, rather than a more direct effect on the program counter that would have led to a greater diversity of target addresses (i.e., one would have observed jumps into the middle of basic blocks). 

The occurrence of such a magic edge may induce an early exit that bypasses all the control blocks of a loop-centric countermeasure. It is not detected by SWIFT either, although it is meant to protect against a very wide range of control-flow hijacking patterns. Overall, this new effect provides a powerful attack vehicle, although it seems difficult at this point to precisely control the target of a magic edge.

\subsection{Fault classification on loops}

Let us now classify all the successful faults observed during the different attack campaigns on loop benchmarks.

As the hardened loops leave some faults undetected, we applied a costly combination of both countermeasures to the \textit{loop2} example from Listing~\ref{loop2-c}. 
We first applied loop scheme from~\cite{plaf} and hardened the resulting code using SWIFT scheme.
We refer to this version as \textit{loop2-sec+swift}. We also ran an attack campaign on this version.

\figurename~\ref{fig:models} shows the distribution of successful faults according to all the fault models established in this paper, for all loop benchmarks. Note that the ratio of instruction skip may be slightly skewed (in excess) due to the priority we assigned to this model in the classification methodology (cf.\ Section~\ref{sec:classification}). Also, the single faults are displayed first for every fault model, before any composite faults.
Note that 15\% of the successful faults remain unexplained. Further investigation is needed, involving additional attack campaigns and analyses.

The results also show that all fault models have been observed in every attack campaign. The proportion of a given fault model varies widely across the test codes, but composite effects are always far from negligible.

\begin{figure}
	\centering
	\begin{tikzpicture}
	\begin{axis}[
	xbar stacked,
	width=19em,
	bar width=9pt,
	y=22pt,
	ymajorgrids = true,
	xmin=0,
	xmax=100,
	ymin=0.2,
	ymax=2.3,
	xtick={0,25,50,75,100},
	xticklabels={0,25,50,75,100\%},
	x tick label style={font=\footnotesize},
	yticklabels={
		loop2-swift+sec (80),
		loop2-swift (447),
		loop2-sec (118),
		loop1-sec (87),
	},
	y tick label style={font=\scriptsize},
	legend style={at={(0.35,-0.6)}, anchor=south, draw=none, legend columns=-1, column sep=0.4ex, font=\footnotesize},
	ytick=data
	]
	\addplot[pattern=north east lines] plot coordinates {
	(42.5, 0.5)
	(01.3, 1)
	(38.1, 1.5)
	(98.0, 2)
};
\addplot[fill=black!40] plot coordinates {
	(07.5, 0.5)
	(34.7, 1)
	(06.8, 1.5)
	(00.0, 2)
};
	\addplot[fill=black!15] plot coordinates {
		(25.0, 0.5)
		(57.7, 1)
		(39.0, 1.5)
		(02.0, 2)
	};
	\addplot[fill=black!65] plot coordinates {
		(25.0, 0.5)
		(06.3, 1)
		(16.1, 1.5)
		(00.0, 2)
	};
	
	\legend{magic edge, other single, mixed, other composite}
	\end{axis}
	
	\end{tikzpicture}
	\caption{Distribution of effects leading to a harmful fault (with the total number of harmful faults for each code)}\label{fig:harmful}
\end{figure}
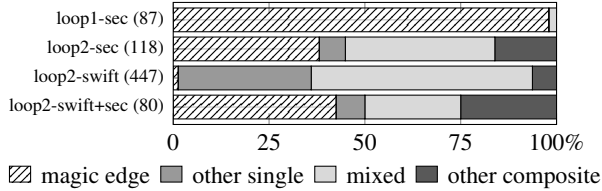

Let us now focus on the classification of the fault effects that manage to bypass the software countermeasures of the hardened loops, by considering only harmful (undetected) faults. \figurename~\ref{fig:harmful} selects some of the effects listed in \figurename~\ref{fig:models}, detailing the breakdown of successful faults.

We can observe that most of the single faults (from \figurename~\ref{fig:models}) are either detected or harmless. The harmful single faults, magic edge excluded, are most often due to an operand substitution (destination register) that leads to a double register corruption that bypasses the detection mechanisms.

Harmful faults are mainly due to composite faults or magic edges.  However, composite faults affecting differently an original computation and its duplicated version are detected: 48\% of the composite faults of \figurename~\ref{fig:models} are no more present in \figurename~\ref{fig:harmful}. The remaining composite faults bypass the countermeasures and represent a large fraction of the undetected faults. 

Magic edges represent a real threat for all the hardened version. A single magic edge cannot be detected by the first (loop-centric) countermeasure. For the hardened versions of \textit{loop2}, around 33\% of mixed faults are a combination of a magic edge and another corruption. In particular, stacking the two countermeasures (version \emph{loop2-swift+sec}) is not sufficient to entirely protect against magic edge and composite effects.

Also, code size may grow exponentially with the stacking of countermeasures: $\times3$ for \textit{loop2-sec} and \textit{loop2-swift} and $\times11$ for \textit{loop2-swift+sec}.
This advocates for further investigation of fault effects on complex microarchitectures to provide a more complete characterization. It would enable the deployment of protection schemes intrinsically resistant to multiple faults. A mix of software and hardware countermeasures seems particularly attractive to mitigate overhead.

\section{Conclusion}\label{sec:ccl}


We first established the vulnerability of loop-based applications to electromagnetic (EM) fault injections targeting an out-of-order superscalar processor. Following a black-box approach and relying on widely available equipment, we demonstrated the exploitation of this vulnerability by locally disrupting the control flow of hardened versions of sensitive loops. We also demonstrated that state-of-the-art software countermeasures could reduce the probability of success of such attacks, but fail to achieve levels of protection comparable with running the same hardened loops on simpler micro-controllers.

We proposed a step-by-step methodology to characterize the fault effects at the level of the instruction set architecture (ISA), starting from extremely simple yet carefully designed code fragments, and following on with loop benchmarks including loops hardened with software countermeasures. We applied this methodology to identify a range of fault models, some already well-known observed on simpler processors such as instruction skip or register corruption, as well as newer ones, potentially very powerful and specific to complex microarchitectures. Among the latter, we observed and classified correlated effects such as \textit{instruction skip and replay}, \textit{operand substitution} and \textit{magic-edge} control flow hijacking, as well as compositions of such effects.

These results explain the wide vulnerabilities left open by state-of-the-art software countermeasures designed for conventional fault models. In particular, they highlight the fundamental weakness of replication-based software-only protections, aimed at single, localized faults. The correlated and mixed fault models we characterize seem to evade such replication tactics. Our results motivate further research in two directions: (1) broadening the characterization of EM fault injection at ISA level, covering microarchitectures from different vendors  taped out with different technological nodes and exploring more die locations, and (2) designing and evaluating hardware or hybrid hardware-software countermeasures capable of detecting multiple fault effects with strong correlations, as well as advanced control flow hijacking.  



\bibliographystyle{abbrv}
\bibliography{archive}

\end{document}